\catcode`\@=11
%
%
\def\psfortextures{
\def\PSspeci@l##1##2{%
\special{illustration ##1\space scaled ##2}%
}}
\def\psfordvitops{
\def\PSspeci@l##1##2{%
\special{dvitops: import ##1\space \the\drawingwd \the\drawinght}%
}}
\def\psfordvips{
\def\PSspeci@l##1##2{%
\d@my=0.1bp \d@mx=\drawingwd \divide\d@mx by\d@my%
\includegraphics{##1\space}%
}}
\def\psforoztex{
\def\PSspeci@l##1##2{%
\special{##1 \space
      ##2 1000 div dup scale
      \putsp@ce{\number-\psllx} \putsp@ce{\number-\pslly} translate
}%
}}
\def\putsp@ce#1{#1 }
\def\psfordvitps{
\def\psdimt@n@sp##1{\d@mx=##1\relax\edef\psn@sp{\number\d@mx}}
\def\PSspeci@l##1##2{%
\special{dvitps: Include0 "psfig.psr"}
\psdimt@n@sp{\drawingwd}
\special{dvitps: Literal "\psn@sp\space"}
\psdimt@n@sp{\drawinght}
\special{dvitps: Literal "\psn@sp\space"}
\psdimt@n@sp{\psllx bp}
\special{dvitps: Literal "\psn@sp\space"}
\psdimt@n@sp{\pslly bp}
\special{dvitps: Literal "\psn@sp\space"}
\psdimt@n@sp{\psurx bp}
\special{dvitps: Literal "\psn@sp\space"}
\psdimt@n@sp{\psury bp}
\special{dvitps: Literal "\psn@sp\space startTexFig\space"}
\special{dvitps: Include1 "##1"}
\special{dvitps: Literal "endTexFig\space"}
}}
\def\psonlyboxes{
\def\PSspeci@l##1##2{%
\at(0cm;0cm){\boxit{\vbox to\drawinght
  {\vss
  \hbox to\drawingwd{\at(0cm;0cm){\hbox{(##1)}}\hss}
  }}}
}%
}
\def\psloc@lerr#1{%
\let\savedPSspeci@l=\PSspeci@l%
\def\PSspeci@l##1##2{%
\at(0cm;0cm){\boxit{\vbox to\drawinght
  {\vss
  \hbox to\drawingwd{\at(0cm;0cm){\hbox{(##1) #1}}\hss}
  }}}
\let\PSspeci@l=\savedPSspeci@l
}%
}
%
%
\newread\psiz@
\newdimen\drawinght\newdimen\drawingwd
\newdimen\psxoffset\newdimen\psyoffset
\newbox\drawingBox
\newif\ifNotB@undingBox
\newhelp\PShelp{Proceed: you'll have a 5cm square blank box instead of
your graphics (Jean Orloff).}
\def\@mpty{}
\def\s@tsize#1 #2 #3 #4\@ndsize{
  \def\psllx{#1}\def\pslly{#2}%
  \def\psurx{#3}\def\psury{#4}
  \ifx\psurx\@mpty\NotB@undingBoxtrue
  \else
    \drawinght=#4bp\advance\drawinght by-#2bp
    \drawingwd=#3bp\advance\drawingwd by-#1bp
  \fi
  }
\def\sc@nline#1:#2\@ndline{\edef\p@rameter{#1}\edef\v@lue{#2}}
\def\g@bblefirstblank#1#2:{\ifx#1 \else#1\fi#2}
\def\psm@keother#1{\catcode`#112\relax}
\def\execute#1{#1}
{\catcode`\%=12
\xdef\B@undingBox{
}  		
\def\ReadPSize#1{
 \edef\PSfilename{#1}
 \openin\psiz@=#1\relax
 \ifeof\psiz@ \errhelp=\PShelp
   \errmessage{I haven't found your postscript file (\PSfilename)}
   \psloc@lerr{was not found}
   \s@tsize 0 0 142 142\@ndsize
   \closein\psiz@
 \else
   \loop
     \execute{\begingroup
       \let\do\psm@keother
       \dospecials
       \catcode`\ =10
       \catcode`\^^M=9
       \global\read\psiz@ to\n@xtline
       \endgroup}
     \ifeof\psiz@
       \errhelp=\PShelp
       \errmessage{(\PSfilename) is not an Encapsulated PostScript File:
           I could not find any \B@undingBox: line.}
       \edef\v@lue{0 0 142 142:}
       \psloc@lerr{is not an EPSFile}
       \NotB@undingBoxfalse
     \else
       \expandafter\sc@nline\n@xtline:\@ndline
       \ifx\p@rameter\B@undingBox\NotB@undingBoxfalse
         \edef\int@rmediateresult{%
           \expandafter\g@bblefirstblank\v@lue\space\space\space}
         \expandafter\s@tsize\int@rmediateresult\@ndsize
       \else\NotB@undingBoxtrue
       \fi
     \fi
   \ifNotB@undingBox\repeat
   \closein\psiz@
 \fi
\message{#1}
}
%
%
\newcount\xscale \newcount\yscale \newdimen\pscm\pscm=1cm
\newdimen\d@mx \newdimen\d@my
\let\ps@nnotation=\relax
\def\psboxto(#1;#2)#3{\vbox{
   \ReadPSize{#3}
   \divide\drawingwd by 1000
   \divide\drawinght by 1000
   \d@mx=#1
   \ifdim\d@mx=0pt\xscale=1000
         \else \xscale=\d@mx \divide \xscale by \drawingwd\fi
   \d@my=#2
   \ifdim\d@my=0pt\yscale=1000
         \else \yscale=\d@my \divide \yscale by \drawinght\fi
   \ifnum\yscale=1000
         \else\ifnum\xscale=1000\xscale=\yscale
                    \else\ifnum\yscale<\xscale\xscale=\yscale\fi
              \fi
   \fi
   \divide \psxoffset by 1000\multiply\psxoffset by \xscale
   \divide \psyoffset by 1000\multiply\psyoffset by \xscale
   \global\divide\pscm by 1000
   \global\multiply\pscm by\xscale
   \multiply\drawingwd by\xscale \multiply\drawinght by\xscale
   \ifdim\d@mx=0pt\d@mx=\drawingwd\fi
   \ifdim\d@my=0pt\d@my=\drawinght\fi
   \message{scaled \the\xscale}
 \hbox to\d@mx{\hss\vbox to\d@my{\vss
   \global\setbox\drawingBox=\hbox to 0pt{\kern\psxoffset\vbox to 0pt{
      \kern-\psyoffset
      \PSspeci@l{\PSfilename}{\the\xscale}
      \vss}\hss\ps@nnotation}
   \global\ht\drawingBox=\the\drawinght
   \global\wd\drawingBox=\the\drawingwd
   \baselineskip=0pt
   \copy\drawingBox
 \vss}\hss}
  \global\psxoffset=0pt
  \global\psyoffset=0pt
  \global\pscm=1cm
  \global\drawingwd=\drawingwd
  \global\drawinght=\drawinght
}}
%
%
\def\psboxscaled#1#2{\vbox{
  \ReadPSize{#2}
  \xscale=#1
  \message{scaled \the\xscale}
  \divide\drawingwd by 1000\multiply\drawingwd by\xscale
  \divide\drawinght by 1000\multiply\drawinght by\xscale
  \divide \psxoffset by 1000\multiply\psxoffset by \xscale
  \divide \psyoffset by 1000\multiply\psyoffset by \xscale
  \global\divide\pscm by 1000
  \global\multiply\pscm by\xscale
  \global\setbox\drawingBox=\hbox to 0pt{\kern\psxoffset\vbox to 0pt{
     \kern-\psyoffset
     \PSspeci@l{\PSfilename}{\the\xscale}
     \vss}\hss\ps@nnotation}
  \global\ht\drawingBox=\the\drawinght
  \global\wd\drawingBox=\the\drawingwd
  \baselineskip=0pt
  \copy\drawingBox
  \global\psxoffset=0pt
  \global\psyoffset=0pt
  \global\pscm=1cm
  \global\drawingwd=\drawingwd
  \global\drawinght=\drawinght
}}
%
\def\psbox#1{\psboxscaled{1000}{#1}}
%
%

%
%
\def\psannotate#1#2{\def\ps@nnotation{#2\global\let\ps@nnotation=\relax}#1}
\def\pscaption#1#2{\vbox{
   \setbox\drawingBox=#1
   \copy\drawingBox
   \vskip\baselineskip
   \vbox{\hsize=\wd\drawingBox\setbox0=\hbox{#2}
     \ifdim\wd0>\hsize
       \noindent\unhbox0\tolerance=5000
    \else\centerline{\box0}
    \fi
}}}

%
\def\at(#1;#2)#3{\setbox0=\hbox{#3}\ht0=0pt\dp0=0pt
  \rlap{\kern#1\vbox to0pt{\kern-#2\box0\vss}}}
%
\newdimen\gridht \newdimen\gridwd
\def\gridfill(#1;#2){
  \setbox0=\hbox to 1\pscm
  {\vrule height1\pscm width.4pt\leaders\hrule\hfill}
  \gridht=#1
  \divide\gridht by \ht0
  \multiply\gridht by \ht0
  \gridwd=#2
  \divide\gridwd by \wd0
  \multiply\gridwd by \wd0
  \advance \gridwd by \wd0
  \vbox to \gridht{\leaders\hbox to\gridwd{\leaders\box0\hfill}\vfill}}
%

%
%
\def\textleftof#1:{
  \setbox1=#1
  \setbox0=\vbox\bgroup
    \advance\hsize by -\wd1 \advance\hsize by -2em}
\def\textrightof#1:{
  \setbox0=#1
  \setbox1=\vbox\bgroup
    \advance\hsize by -\wd0 \advance\hsize by -2em}
\def\endtext{
  \egroup
  \hbox to \hsize{\valign{\vfil##\vfil\cr%
\box0\cr%
\noalign{\hss}\box1\cr}}}
%
\def\frameit#1#2#3{\hbox{\vrule width#1\vbox{
  \hrule height#1\vskip#2\hbox{\hskip#2\vbox{#3}\hskip#2}%
        \vskip#2\hrule height#1}\vrule width#1}}
\def\boxit#1{\frameit{0.4pt}{0pt}{#1}}
\catcode`\@=12 
%
 \psfordvips   
%
\input phyzzx.tex
%
%
\def\refpunct#1#2{\rlap#2\refmark{#1}}
\def\RREF#1#2{\gdef#1{\REF#1{#2}#1}}
\def\CERN{\address{Theory Division, CERN\break
          CH-1211 Geneva 23, Switzerland}}
\def\ULB{\address{Service de Physique Th\'eorique \break
Universit\'e Libre de Bruxelles, Boulevard du Triomphe \break
      CP 225, B-1050 Bruxelles, Belgium}}
%
\def\PRL{ {\sl Phys. Rev. Lett.}   }
\def\PR { {\sl Phys. Rev.  }       }
\def\NP { {\sl Nucl. Phys.}        }
\def\ZP { {\sl Z. Phys. }          }

\def\PH { {\sl Physica}		}
\RREF\weinberg{S. Weinberg, \PH {\bf A96} (1979) 327; H. Georgi,
   		{\it Weak Interactions and Modern Particle Theory}
              Benjamin-Cummings, Menlo Park, CA (1984).}
\RREF\efflag{C.J.C.~Burges and H.J.~Schnitzer, \NP {\bf B228}
         (1983) 464;
          C.N.~Leung, S.T.~Love  and S.~Rao, \ZP {\bf C31} (1986) 433;
          W. Buchm\"uller and D. Wyler, \NP {\bf B268} (1986) 621.   }

\RREF\deRuja{A. de R\'ujula, B. Gavela, P. Hern\'andez and E.~Mass\'o,
        preprint CERN-TH.6272/91 (1991). }
\RREF\deRujb{A. de R\'ujula, B. Gavela, O. P\`ene and F.J. Vegas,
	\NP {\bf B357 } (1990) 311.}
\RREF\Einhorn{M.B. Einhorn and J. Wudka, Santa-Barbara
    preprint NSF--ITP--92--01 Preprint
    (1992).}
\RREF\WW{C.L. Bilchak and J.D. Stroughair, \PRL {\bf D30}
        (1984) 1881.}
\RREF\Carter{J. Carter, eedings of the
        {\sl Proc. Joint International  Lepton--Photon Symposium \&
        Europhysics Conference on High Energy Physics}, Geneva,
1991, eds. S. Hegarty et al., (World Scientific, Singapore, 1992).
} \RREF\AHMQ{F. del Aguila, W. Hollik, J.M. Moreno and M. Quir\'os, \NP
           {\bf 372} (1992) 3. }
\RREF\ZNLC{A. Djouadi, A. Leike, T. Riemann, D. Schaile and
      C. Verzegnassi, preprint CERN-TH.6350/91 (1991)}
\RREF\Robi{R.W. Robinett and J.L. Rosner, \PR {\bf D25} (1982) 3056.}
\RREF\radcorr{M.~Bohm et al., {\sl Nucl. Phys.} {\bf B304} (1988) 463.}
\TABLE\?{Matter content and charges assignment of the extended
model $SU(2)\times
          U(1)\times U(1)^\prime$.}

\TABLE\?{$A_{135}$-values for the SM and for $SU(2) \times
U(1) \times U(1)'$ with $\varepsilon = 0, \pm 0.2$. $A_{135}$ is defined
as follows:
$$A_{135} = {\sigma_{\theta >135} - \sigma_{\theta < 135}
\over \sigma_{total}}.
$$}

\FIGURE\?{Detail of the unpolarized differential cross-section for
the process $e^+e^- \rightarrow W^+W^-$ at $\sqrt s= 200$~GeV: the
solid line shows the SM, the dashed one is for
$\varepsilon= 0.2$ and the dash-dotted one for $\varepsilon =-0.2$;
$\theta$ is the angle between $e^-$ and  $W^+$.}

\FIGURE\?{Same as fig.1 at $\sqrt s =260$ GeV.}
\nopubblock
\titlepage
\line{\hfil\vbox{
\hbox{CERN--TH.6573/92}
\hbox{ULB--TH--04/92}
\hbox{hepth@xxx/9207258}
\hbox{July 1992}
}}

\title{LEP1 {\it vs.} Future Colliders:\break
       Effective Operators And Extended Gauge Group}

\author{J.-M. Fr\`ere\footnote{\dag}{Ma\^{\i}tre
de Recherche FNRS.}, J.M. Moreno\footnote{*}{Supported in part
by IISN.}, M. Tytgat\footnote{\flat}{Aspirant FNRS.}} \ULB
\andauthor{J. Orloff\footnote{\sharp}{e-mail: orloff@dxcern.cern.ch}}
\CERN
\abstract
{In an effective Lagrangian approach to physics beyond the Standard
Model, it has been argued that imposing $SU(2) \times U(1)$
invariance  severely restricts the discovery potential of future
colliders. We exhibit a possible way out in an extended gauge group
context.}   \endpage

\section{INTRODUCTION}

In studying departures from the Standard Model (SM),  effective
Lagran\-gians\refmark{\weinberg,\efflag} prove a convenient
tool,  since they unify the various manifestations of a given
contribution. Usually an operator expansion is used and the various
possible operators are ranked by increasing dimensionality.  To any
given order, only a finite number of operators need be considered,
and this is true irrespective of the nature of the underlying SM
extension (which only determines the coefficients).

Gauge invariance imposed on an effective Lagrangian and precise
measurements at LEP1 have recently been shown \refmark{\deRujb}
\refmark{\deRuja} \refmark{\Einhorn} to severely restrict the possible
impact of new physics on observations at forthcoming colliders (LEP2,
...). The argument, with which we are in complete agreement, proceeds
in two steps:
\item{\bullet} In the framework of gauge theories, extra terms induced
by new physics in an effective Lagrangian must at least obey the
$SU(2)_L \times U(1)$ invariance.  In particular, they must be
expressible (above the scale of symmetry breakdown) in terms of
$[SU(2)_L \times U(1)]$-invariant products of fields, including the
SM scalar doublet(s).
\item{\bullet} These operators can be classified by increasing
dimensionality.  If the new physics scale is high enough, only the
lowest-dimensional operators will contribute.

Gauge invariance thus strongly limits the number of operators available
for a given dimensionality; consequently, the typical effects expected
at future colliders, such as new 3-vector vertices, are intimately
related to modifications of typical LEP1 observables (widths,
asymmetries, ...).  When the analysis is limited to dimension 6
operators, and apart from a few  blind directions, the LEP1
measurements\refmark{\Carter} tend to give stronger bounds on the
coefficients of these operators than those expected from the direct
observation of the 3-vector vertices.

We found it interesting to explore some limitations of this approach.
Our goal was to see how the above predictions would be affected when
the implicit assumption of a very high mass scale for any new
physics was relaxed.

We have chosen to study a specific example, built upon the extended
group $SU(2) \times U(1) \times U(1)'$\refmark{\Robi}.  This introduces
essentially two new fields, namely an extra gauge boson
$(B^\prime_{\mu})$ and an extra scalar singlet $(\chi)$ (an extra
$\nu_R$ per generation is also  needed, so as to avoid anomalies, but
it plays no role in the present discussion). The mass of the extra gauge
boson will characterize the scale of this part of the new physics. If
this scale lies far beyond the one accessible at future colliders, the
extra degrees of freedom may be integrated out in an effective
lagrangian approach, which leads, for dimension-6 operators, to the
same analysis as {Ref.}[\deRuja].

We have instead assumed the mass of the new gauge gauge boson to be
relatively close to the LEP2 scale.

One systematic way to treat this model would be to consider the higher
orders in a dimensional expansion.  This, however, quickly becomes
intractable---in view of the number of new operators involved---and
unreliable, since doubts arise about the convergence of the expansion
(specially close to a resonance); it is then safer to bring in the full
gauge group structure $SU(2) \times U(1) \times U(1)'$.

In the spirit of Ref.[\deRuja], we continue to use an effective
Lagrangian approach (limited to dimension-6 operators) to parametrize
any new physics beyond the $Z^\prime$ ({\frenchspacing i.e.} the
heavy-mass eigenstate). The only difference lies in the fact that the
effective operators  are now to be classified following their
invariance under the gauge group $SU(2) \times U(1) \times U(1)'$
rather than $SU(2) \times U(1)$.

At first sight, one might expect that extending the requested symmetry
group would further restrict the choice of operators.  It turns out
that this is more than compensated for by the presence of the new
degrees of freedom $Z^\prime$ and $\chi$.

A typical example involves couplings of $W$ bosons.  In the
approach of {Ref.}[\deRuja], these can only arise from two dimension-6
operators, namely:
$$\eqalignno{ O_{WB} &= \phi^+ \sigma^i \phi W_{\mu
\nu}^i B^{\mu \nu}
  &\eqname\eqi \cr
O_W &= {1 \over 3 !} \epsilon_{ijk} W_\mu^{\ \nu j} W_\nu^{\ \lambda
  k} W_\lambda^{\ \mu i}.
  &\eqname\eqii \cr}$$
In the present case, we can obviously add to this list the operator:
$$O_{W B^\prime} = \phi^+ \sigma^i \phi W_{\mu \nu}^i B^{\prime \mu
\nu}. \eqn\eqiii
$$
Such an operator is absent from the expansion of {Ref.}[\deRuja].
Nevertheless, (\eqiii) will generate anomalous couplings between the
physical $Z$ (now a mixture of $W^3, B, B^\prime$) and $W$'s.

One needs to formally integrate out $B^\prime$ (and $\chi$) to compare
the two approaches.  The situation is most easily exemplified in terms
of graphs.  In particular, for $Z W^+ W^-$  coupling, we get
contributions from :
$$
\psannotate{\psbox{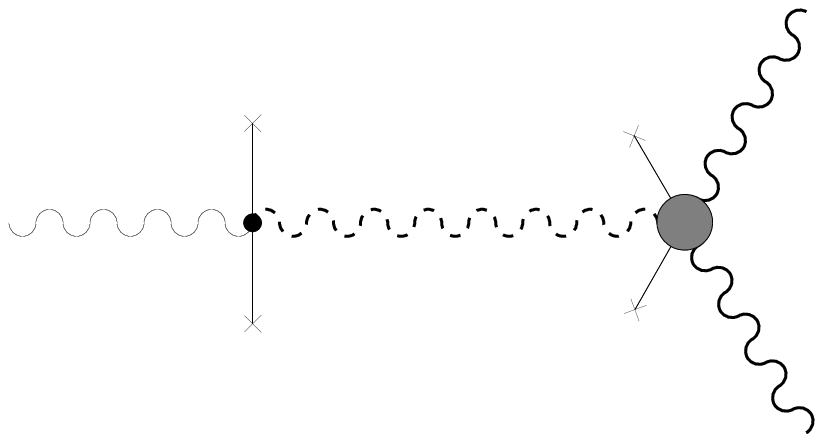}}
{
\at(0.2\pscm;2.5\pscm){$B$}%
\at(4.5\pscm;2.5\pscm){$B'$}%
\at(2.4\pscm;0.7\pscm){$\phi$}%
\at(2.4\pscm;3.5\pscm){$\phi^\dagger$}%
\at(6\pscm;0.7\pscm){$\phi$}%
\at(6\pscm;3.5\pscm){$\phi^\dagger$}%
\at(8.2\pscm;3.8\pscm){$W^+$}%
\at(8.2\pscm;0.5\pscm){$W^-$}%
}
$$
Upon integrating out $B^\prime$ and $\chi$ (the presence of the $\chi$
field is implicit in the $B^\prime$ mass term), we get:
$$
\psannotate{\psbox{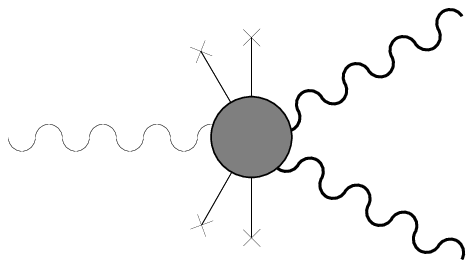}}
{
\at(0\pscm;1.5\pscm){$B$}%
\at(2.1\pscm;0\pscm){$\phi$}%
\at(2.7\pscm;0\pscm){$\phi$}%
\at(2.1\pscm;2.5\pscm){$\phi^\dagger$}%
\at(2.7\pscm;2.5\pscm){$\phi^\dagger$}%
\at(4.5\pscm;1.8\pscm){$W^+$}%
\at(4.5\pscm;0.5\pscm){$W^-$}%
}
$$
containing {e.g.} the new operator (of dimension 8 in the fields)
$${1 \over M_{B^\prime}^2} \phi^+\phi\ B_{\mu \nu}\ \phi^+ \tau^j
\phi\  W^{j \mu \nu}. \eqn\eqiv
$$
On top of other dimension-8 operators, higher dimensionalities will of
course appear through the low-momentum expansion of the $B'$
propagator.

Let us now turn back to the full $SU(2) \times U(1) \times U(1)'$
and review briefly the parameters involved (we use $g_2, g_1,
g_1^\prime$ for the respective gauge couplings and $Y, Y^\prime$ for
the hypercharges $v=\sqrt{2}|\vev{\phi}|,\  V =|\vev{\chi}|$).  The
field content and quantum number assignments are detailed in Table 1.

We now include the additional operator (3) in the Lagrangian
density:
$$
{\cal L}_{W B^\prime} = {\varepsilon \over v^2} . O_{W
B^\prime}.\eqn\eqviii
$$
We have followed {Ref.}[\deRuja] in the definition  of $\varepsilon$.

In addition to 3-gauge boson couplings of interest for future
colliders, this implies (after symmetry breaking) corrections to the
kinetic energy of neutral gauge bosons. The $Z$-propagator is thus
modified, with potential effects at LEP1. It is however easy to check
that the propagator of the light $Z$ physical state is not directly
affected by the $\varepsilon$ correction, but only by the combination
$\varepsilon  \theta_{3}$, where $\theta_3$ is the standard $Z$-$Z'$
mixing angle. As a result, the departure from a pure $SU(2) \times U(1)
\times U(1)'$ is kept minimal once $\theta_{3}$ remains within its
usual bounds \refmark{\AHMQ}.

The same cannot be said of the $e^+ e^- \to W^+ W^-$ cross section.
The effective operator (3) manifests itself in two ways in this
process:
\item{\bullet}The currents are modified by the redefinition of the
boson fields ({\it e.g.} the $B'$ content of $Z$)
\item{\bullet}Direct contributions to the three boson vertices appear.
In particular, the $Z' - W^+ W^-$  ($O(\theta_3)$) vertex already
present in the extended $SU(2) \times U(1) \times U(1)'$ now
receives an $O(\varepsilon)$ contribution. That coupling is not
suppressed by $\theta_3$ and thus becomes competitive with the other
contributions, with which it interferes.  Its angular dependence makes
it quite conspicuous.

\section{OVERVIEW OF THE CALCULATIONS}

The differential cross-section for the process $e^+ e^- \rightarrow
W^+ W^-$ (see {Ref.}[\WW]) is
$${ {d \sigma (e^+e^- \rightarrow W^+W^-)} \over {d \cos \theta} }
    = s^{1/2}(s/4 - M_W^2)^{1/2}  {\vert A \vert^2},  \eqn\eqaxvi
$$
where $A$ takes into account the contributions from the four usual
diagrams corresponding to the t-channel $\nu$ exchange and the
s-channel $\gamma, Z, Z'$ exchanges
$$
\vert A \vert^2 = {1 \over 8 \pi}
    \sum_{\alpha,\beta} \left(
   a_{\alpha}^V\ Spin_{\alpha ,\beta}\ a_{\beta}^V +
   a_{\alpha}^A\ Spin_{\alpha ,\beta}\ a_{\beta}^A
                         \right),
  \eqn\eqaxvii
$$
with
$$
a^{V(A)}_{\alpha = \nu,\gamma, Z, Z'} =
\left( {  g^2_2  \over 4 t  },
       { e^\gamma_{V(A)} g^\gamma \over s},
       { e^Z_{V(A)} g^Z       \over s - M_Z^2},
       { e^{Z'}_{V(A)} g^{Z'} \over s - M_{Z'}^2} \right)\eqn\eqaxviii
$$
and
$$ \eqalign{
Spin_{\nu,\nu }
 &=\left[{ut \over {M_W^4} } -1 \right]
   \left[{t^2 \over 4 s^2} +{M_W^4 \over s^2}\right]
   + {t^2\over s M_W^2}\cr
Spin_{\nu,i}&=\left[{ut \over {M_W^4} } - 1 \right]
 \left[{ \kappa_i t\over 4 s} - {M_W^2 t\over 2s^2}
  - {M_W^4 \over s^2}\right]
  + (1+\kappa_i)\left[{ t \over 2M_W^2}
  - {t\over s} + {M_W^2 \over s}
 \right]\cr
Spin_{i,j}&=\left[{ut \over {M_W^4} } - 1 \right]
 \left[{ \kappa_i\kappa_j \over 4}
  - {M_W^2 \over s}{(1 + \kappa_i\kappa_j) \over 2}
  + 3{M_W^4 \over s^2}
 \right]\cr
&\ \ + (1+\kappa_i)(1+\kappa_j) \left[{s \over 4M_W^2}-1\right]
    ;\hskip2cm i=\gamma, Z, Z'\cr}
\eqn\eqaxix $$
describing the angular dependence. The different couplings
$e^i_{V(A)}$, $g^i$, $\kappa^i$ are defined through the matrix $\cal
S$ relating the physical neutral fields to the original gauge
fields\foot{{\cal S} is {\em not} unitary, as
$\varepsilon_{WB'}$ induces a non-canonical kinetic term requiring a
rescaling of the fields.}:
$$
\pmatrix{ W_\mu^3 \cr  B_\mu \cr  B'_\mu \cr}
 = {\cal S}.\pmatrix {A_\mu \cr  Z_\mu \cr  Z'_\mu \cr}\eqn\eqaxvx
$$
by
$$\eqalign{
(g^\gamma,\, g^Z, \,g^{Z'})
 & =(g_2,0,0) . {\cal S}\cr
\left( g^\gamma(1+\kappa^\gamma), \,g^Z(1+\kappa^Z),
\,g^{Z'}(1+\kappa^{Z'})\right)
 & = (g_2, \,0,\,\varepsilon) .{\cal S}\cr
\left( e^\gamma_{L(R)},\,e^Z_{L(R)},\,e^{Z'}_{L(R)}\right)
 & =(g_2 T^3_{e_{L(R)}}, \, g_1 Y^B_{e_{L(R)}}, \,g'_1
      Y^{B'}_{e_{L(R)}}).{\cal S}\cr
e_{V(A)}
 &  = {1 \over 2}(e_R + (-) e_L). \cr }\eqn\eqaxvxi
$$

\section{DISCUSSION OF THE NUMERICAL RESULTS}

We have not pursued a systematic search of the parameter space, but
present here a simple example. While the parameters are obviously
chosen to make our point clear, we have not attempted to maximize the
effects to the extreme limits allowed by current data. We have taken a
smallish value for the new coupling strength $g_1'$ ($\lambda = g_1'/
g_2 = 0.1$), which allows for a relatively small $Z'$-mass
($M_{Z'} = 300$  GeV) without dangerous direct contributions. In the
same spirit, the  $Z$-$Z'$ mixing is kept small ($\theta_3 \sim 0.008$)
to control the indirect ones\refmark{\AHMQ}.

As we mentioned above, all the effects of the additional operator
$O_{WB'}$ at LEP1 are suppressed by a factor $\theta_3 \varepsilon$. We
do not present here complete fits to the LEP1 data, but use instead the
evolution of $\sin^2 \theta_W \vert_{eff}$ as the dominant
constraint\foot{One can check that the same na{\"\i}ve approach applied
to the parameter $\varepsilon_{WB}$ of {Ref.}[\deRuja] essentially
reproduces their global fit.}.  Taking values of $\varepsilon$ between
$-0.2$ and $0.2$, we find $\Delta \sin^2 \theta_W \vert_{eff} \leq
0.001 \sim 1 \sigma $ \refmark{\Carter}. The fact that the values for
$\varepsilon$ are larger than those found for the similar parameter
$\varepsilon_{WB} $\refmark{\deRuja} is just an illustration of the
screening effect due to the small mixing angle $\theta_3$.

We have plotted the differential cross-section at 200 GeV, which
corresponds approximately to the maximum of the  $WW$ cross-section.
We also give a plot for 260 GeV, to show how the effect of $\varepsilon$
is boosted by nudging the energy only a little closer to the pole of
the $Z'$.

We first make out some qualitative remarks. With $\varepsilon$ set to
$0$, the cross-section is indistinguishable from the SM one. It is an
interesting question to know whether other channels might reveal the
presence of the extended gauge group\refmark{\ZNLC} independently of
gauge boson couplings. We have only checked the most obvious
channel---$\mu$-pair production---and found that while the relative
difference in cross-section was indeed sizeable, the
 overall value of this cross-section was very small, which might
cause a problem with statistics.

The presence of the anomalous gauge boson couplings controlled by
$\varepsilon$ reflects in a modification of the cross-section. This is
not uniform, and a decrease in the backward part is compensated for by
an increase in the forward one. A detailed study of the best
observables to detect the effect depends obviously on the detailed
properties of the detectors (angular resolution, ...) and falls
beyond the scope of this paper.  If one wants to use a simple number
to quantify the departures from the SM, inspection of the curves
suggests resorting to $A_{135} = {\sigma_{\theta >135} -
\sigma_{\theta < 135} \over
\sigma_{total}}$ (the angle 135 is close to optimal at 200
GeV, and should be adapted as a function of energy) or to the deviation
at the maximum of the cross-section.

The numerical results are displayed in figs. 1 and 2 and gathered in
table 2 for two values of the centre-of-mass energy ({e.g.} 200 and
260 GeV). As can be expected, the effects become more important when
approaching the $Z'$ pole. We have not taken into account the
radiative corrections for $WW$ production (but included them for LEP1)
as we do not expect them to qualitatively alter our conclusions.
Indeed they were shown in the Standard Model to be almost
$\theta$-independent\refpunct\radcorr, depleting at most the
irrelevant small $\theta$ region by 15\%. Altough the presence of the
$Z'$ will slightly change their behaviour, this can be considered as a
second-order correction for our discussion.

As we mentioned above, we have not done an exhaustive exploration of
the parameter space, since our purpose was rather to illustrate our
proposition than to give a full account of the effects of this kind of
operators in the context of extended gauge groups.

\section{CONCLUSIONS}

The model we have examined here shows one possible way to evade the
limits of {Ref.}[\deRuja]. It also gives some idea of the price to
pay to achieve this goal. It requires both a relatively light
$Z^\prime$ and anomalous couplings of that $Z^\prime$ (themselves
attributed to unspecified new physics) which are quite sizeable.

What can we conclude from the above approach?
\item{\bullet} Despite strong constraints arising from the
(high-luminosity) precise measurements at LEP1 and lower energies, the
introduction of a larger gauge group, broken at a scale higher but still
comparable with the SM, considerably increases the allowed freedom.

\item{\bullet} In particular, the {\it anomalous} couplings of an extra
$Z$ boson are not considerably restricted at the LEP1 level, and may
lead to important departures from SM expectations at energies
reachable in the near future.

The present observations do not detract from the importance of
{Ref.}[\deRuja], but stress the point that $e^+ e^-$ colliders should be
designed with enough flexibility to be operated as discovery machines.

\subsection{Acknowledgements}

We wish to thank Alain Blondel, Andy Cohen, Alvaro de R\'ujula,
Bel\'en Gavela, Gordy Kane, Olivier P\`ene and Mariano Quir\'os.

\refout
\tabout
\endpage
\vbox {\offinterlineskip
\hrule
\halign
{&\vrule#&\strut\quad\hfil#&\vrule#&\strut\quad
\hfil#\quad&\vrule#&\strut\quad\hfil#\quad
&\vrule#&\strut\quad\hfil#\quad&\vrule#&\strut\quad
\hfil#\quad&\vrule#&\strut\quad\hfil#\quad&\vrule#
&\strut\quad\hfil#\quad&\vrule#&\strut\quad\hfil#\quad\cr
height2pt&\omit&&\omit&&\omit&&\omit&&\omit&&\omit&&\omit&&\omit&&\omit&\cr
& Hypercharge \hfil && ${\pmatrix{u\cr  d\cr}}_L$ && $u_L^c$ &&
$d_L^c$ && ${\pmatrix{\nu\cr e\cr}}_L$ && $e_L^c$ && $\nu_L^c$ &&
$\pmatrix{\phi_1\cr \phi_2\cr}$ && $\chi$\ \ &\cr
height2pt&\omit&&\omit&&\omit&&\omit&&\omit&&\omit&&\omit&&\omit&&\omit&\cr
\noalign{\hrule}
height2pt&\omit&&\omit&&\omit&&\omit&&\omit&&\omit&&\omit&&\omit&&\omit&\cr
 & \ \ && \ \ && \ \ && \ \ && \ \ && \ \ && \ \ && \ \ && \ \ & \cr &
$Y$ \ \ \ \ \ \ && ${1\over 6}$ \ \ \ \ && $ -{2\over 3}$ && ${1\over
3}$ && $-{1\over 2}$ \ \  && $\scriptstyle1$ && $\scriptstyle 0$ &&
${1\over 2}$ \ \ && $\scriptstyle 0$\ \ &\cr & \ \ && \ \ && \ \ && \
\ && \ \ && \ \ && \ \ && \ \ && \ \ & \cr
& $Y'$ \ \ \ \ \ \ &&
${1\over5}$ \ \ \ \ && $ {1\over 5}$ && $ -{3\over 5}$ && $ -{3\over
5}$ \ \  &&  ${1\over5}$ && $\scriptstyle1$ && $-{2\over 5}$ \ \ &&
$\scriptstyle1$\ \ &\cr & \ \ && \ \ && \ \ && \ \ && \ \ && \ \ && \
\ && \ \ && \ \ & \cr height2pt
&\omit&&\omit&&\omit&&\omit&&\omit&&\omit&&\omit&&\omit&&\omit&\cr}
\hrule}
\vskip 0.5in
\centerline  {Table 1}

\vskip 2in

\hskip 1in
\vbox {\offinterlineskip
\hrule
\halign
{&\vrule#&\strut\quad#\hfil\quad&\vrule#&\strut\quad\hfil#\quad\cr
height2pt&\omit&&\omit&&\omit&\cr & Model \hfil  && $\sqrt s = 200$
GeV && $\sqrt s = 260$ GeV &\cr height2pt&\omit&&\omit&&\omit&\cr
\noalign{\hrule} height2pt&\omit&&\omit&&\omit&\cr & \ \ && \ \ && \ \
&\cr & SM &&  $-0.064$ \ \ \ \ && $0.297$ \ \ \ \ &\cr & \ \  && \ \
&& \ \ &\cr & $\varepsilon = 0$  && $-0.066$ \ \ \ \  && $0.292$ \ \ \
\  &\cr & \ \  && \ \ && \ \ &\cr & $\varepsilon = + 0.2$  && $-0.061$
\ \ \ \  && $0.290$ \ \ \ \  &\cr & \ \ &&  \ \ && \ \ &\cr &
$\varepsilon = - 0.2$  && $-0.089$ \ \ \ \  && $0.180$ \ \ \ \  &\cr &
\ \ &&  \ \ && \ \ &\cr height2pt&\omit&&\omit&&\omit&\cr}
\hrule}
\vskip 0.5in
\centerline  {Table 2}
\endpage
\figout
\endpage
\psboxto(\hsize;\vsize){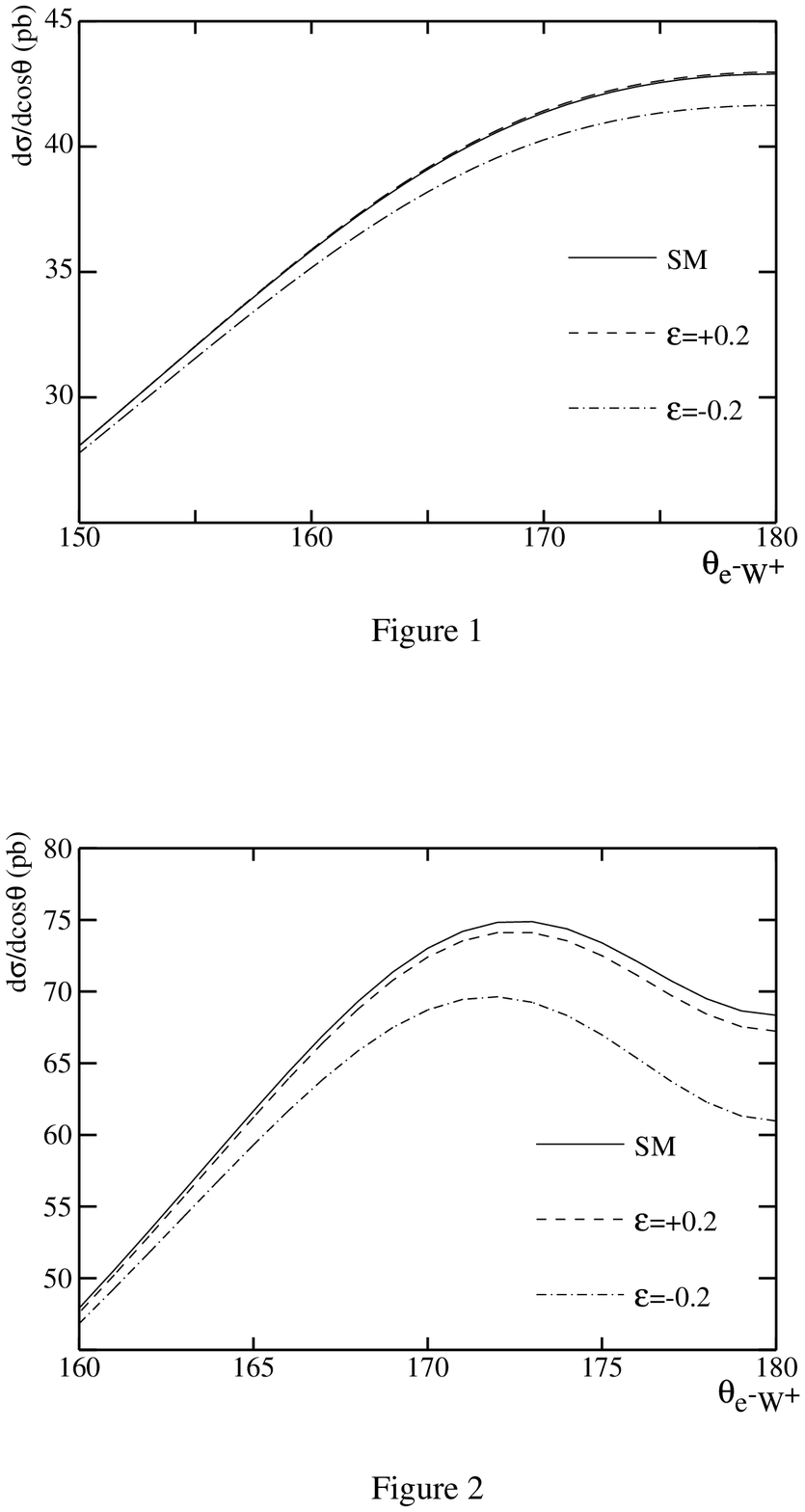}
\bye